\documentclass[a4paper, amsfonts, amssymb, amsmath, reprint, showkeys, nofootinbib, twoside,superscriptaddress]{revtex4-1}
\usepackage[english]{babel}
\usepackage[utf8]{inputenc}

\usepackage{amsthm}
\usepackage{mathtools}
\usepackage{hyperref}
\usepackage{natbib}
\usepackage{physics}
\usepackage{xcolor}
\usepackage{graphicx}
\usepackage[left=23mm,right=13mm,top=35mm,columnsep=15pt]{geometry} 
\usepackage{adjustbox}
\usepackage{placeins}
\usepackage[T1]{fontenc}
\usepackage{csquotes}
\usepackage{xcolor}

\definecolor{acolour}{HTML}{E66101}
\definecolor{bcolour}{HTML}{B2ABD2}
\definecolor{ccolour}{HTML}{5E3C99}

\definecolor{linkcolor}{HTML}{5E3C99}
\definecolor{urlcolor}{HTML}{D7191C}

\hypersetup{
    pdfusetitle,
    unicode=true,          % non-Latin characters in Acrobat’s bookmarks
    pdftoolbar=true,        % show Acrobat’s toolbar?
    pdfmenubar=true,        % show Acrobat’s menu?
    pdffitwindow=false,     % window fit to page when opened
    pdfstartview={FitH},    % fits the width of the page to the window
    pdfsubject={Communications library},   % subject of the document
    pdfkeywords={robust communications} {RabbitMQ} {workflows} {high-throughput} {big-data} {computational science} {AiiDA}, % list of keywords
    pdfnewwindow=true,      % links in new PDF window
    colorlinks=true,
    linkcolor=linkcolor,          % color of internal links (change box color with linkbordercolor)
    citecolor=acolour,        % color of links to bibliography
    filecolor=magenta,      % color of file links
    urlcolor=urlcolor           % color of external links
}

\begin{document}
\title{kiwiPy: Robust, high-volume, messaging for big-data and computational science workflows}

\author{Martin Uhrin}
    \email[Correspondence email address: ]{martin.uhrin.10@ucl.ac.uk}% Your name
    \affiliation{Department of Energy Conversion and Storage, Technical University of Denmark, 2800 Kgs. Lyngby, Denmark}
    \affiliation{National Centre for Computational Design and Discovery of Novel Materials (MARVEL), École Polytechnique Fédérale de Lausanne, CH-1015 Lausanne, Switzerland}
    \affiliation{Theory and Simulation of Materials (THEOS), Faculté des Sciences et Techniques de l’Ingénieur, École Polytechnique Fédérale de Lausanne, CH-1015 Lausanne, Switzerland}
    
\author{Sebastiaan P. Huber}
    \affiliation{National Centre for Computational Design and Discovery of Novel Materials (MARVEL), École Polytechnique Fédérale de Lausanne, CH-1015 Lausanne, Switzerland}
    \affiliation{Theory and Simulation of Materials (THEOS), Faculté des Sciences et Techniques de l’Ingénieur, École Polytechnique Fédérale de Lausanne, CH-1015 Lausanne, Switzerland}

\date{\today} % Leave empty to omit a date

\begin{abstract}
In this work we present kiwiPy, a Python library designed to support robust message based communication for high-throughput, big-data, applications while being general enough to be useful wherever high-volumes of messages need to be communicated in a predictable manner.  KiwiPy relies on the RabbitMQ protocol, an industry standard message broker, while providing a simple and intuitive interface that can be used in both multithreaded and coroutine based applications.  To demonstrate some of kiwiPy's functionality we give examples from AiiDA, a high-throughput simulation platform, where kiwiPy is used as a key component of the workflow engine.
\end{abstract}


\maketitle

\section{Summary}

The computational sciences have seen a huge increase in the use of high-throughput, automated, workflows over the course of the last two decades or so.  Focusing on just our domain of computational materials science, there have been several large scale initiatives to provide high-quality results from standardised calculations \citep{Landis2012, Curtarolo2012, Jain2013, Saal2013, Draxl2019a, Talirz2020}.  Almost all of these repositories are populated using results from high-throughput quantum mechanical calculations that rely on workflow frameworks \citep{Jain2015a, Mortensen2020}, including our own AiiDA\footnote{\url{http://www.aiida.net/}}\citep{Pizzi2016, Huber2020} which powers the Materials Cloud\footnote{\url{https://www.materialscloud.org/}}.  One of the many challenges for such frameworks is maximising fault-tolerance whilst simultaneously maintaining high-throughput, often across several systems (typically the client launching the tasks, the supercomputer carrying out the computations and the server hosting the database).  

On the software level these problems are perhaps best addressed by using messaging brokers that take responsibility for guaranteeing the durability and atomicity of messages and often enable event based communication.  Indeed, solutions such as RabbitMQ\footnote{\url{https://www.rabbitmq.com/}} see widespread adoption in industry, however, adoption in academia has been more limited, with home-made queue data structures, race condition susceptible locks and polling based solutions being commonplace. This is likely due to message brokers typically having complex APIs (which reflect the non-trivial nature of the underlying protocol) as well as the lack of familiarity with event-based programming in general within the community.  KiwiPy\footnote{\url{https://kiwipy.readthedocs.io/en/latest/}} was designed specifically to address both these issues, by providing a tool that allows building robust event based systems with an interface that is as simple as possible.

KiwiPy provides three main message types to the user, task queues, Remote Procedure Calls (RPCs), and, broadcasts. These are all exposed via one class called the `Communicator' which can be trivially constructed by providing a URI string pointing to the RabbitMQ server. This starkly contrasts with the several steps that are required to establish a connection to perform each of the message types listed above using standard libraries\footnote{\url{https://www.rabbitmq.com/getstarted.html}}. Furthermore, by default, kiwiPy creates a separate communication thread that the user never sees, allowing them to interact with the communicator using familiar Python syntax, without the need to be familiar with either coroutines or multithreading.  This has the additional advantage that kiwiPy will maintain heartbeats (a periodic check to make sure the connection is still alive) with the server whilst the user code can be doing other things. Heartbeats are an essential part of RabbitMQ's fault tolerance whereby two missed checks will automatically trigger the message to be requeued to be picked up by another client.

To demonstrate some of the possible usage scenarios, we briefly outline the way kiwiPy is used in AiiDA.  AiiDA, amongst other things, manages the execution of complex workflows made up of processes which may have checkpoints.

\subsection{Task queues}

As is common for high-throughput workflow engines, AiiDA maintains a task queue to which processes are submitted (typically from the user's workstation).  These tasks are then consumed by multiple daemon processes (which may also be on the user's workstation or remote) and will only be removed from the task queue once they have been acknowledged to be completed by the consumer.  The daemon can be gracefully or abruptly shut down and no task will be lost, since the task will simply be requeued by the broker once it notices that the consumer has died.  Furthermore, there are no worries of race conditions between multiple daemon processes, since the task queue guarantees to only distribute each task to, at most, one consumer at a time.

\subsection{Remote Procedure Calls}

These are used to control live processes.  Each process has a unique identifier and can be sent a `pause', `play' or `kill' message, the response to which is optionally sent back to the initiator to indicate success or otherwise.     

\subsection{Broadcasts}

These currently serve two purposes: sending `pause', `play' or `kill' messages to all processes at once by broadcasting the relevant message and to control the flow between processes.  If a parent process is waiting for a child to complete it will be informed of this via a broadcast message by the child saying that its execution has terminated. This enables decoupling as the child need not know about the existence of the parent.

Together these three message types allow AiiDA to implement a highly-decoupled, distributed, yet, reactive system that has proven to be scalable from individual laptops to workstations, driving simulations on high-performance supercomputers with workflows consisting of varying durations, ranging from milliseconds up to multiple days or weeks.

It is our hope that by lowering the barriers to adoption, kiwiPy will bring the benefits of industry grade message brokers to academia and beyond, ultimately making robust scientific software easier to write and maintain.  

\section{Acknowledgements}

We would like to thank Giovanni Pizzi, Nicola Marzari, and the AiiDA team for their continuous coordination and development of the project.  We also thank Jason Yu for contributing the first version of the documentation.
This work is supported by the MARVEL National Centre for Competency in Research funded by the Swiss National Science Foundation (grant agreement ID 51NF40-182892) and the European Materials Modelling Council-CSA funded by the European Union‘s Horizon 2020 research and innovation programme under Grant Agreement No 723867.

\bibliography{ms}

\end{document}